\documentclass[12pt]{article}
\usepackage{latexsym,amssymb}
\newtheorem{definizione}{Definition}[section]
\newcommand{\bdefi}{\begin{definizione}}
\newcommand{\edefi}{\end{definizione}}
\makeatletter
\@addtoreset{equation}{section}
\makeatother

\newcommand{\ft}[2]{{\textstyle\frac{#1}{#2}}}
\def\tilde{\widetilde}
\def\l{\lambda}
\def\1bar{1\hskip -.275cm -}
\def\2bar{2\hskip -.275cm -}
\def\3bar{3\hskip -.275cm -}
\def\n010{N^{0,1,0}}
\newcommand{\eqn}[1]{(\ref{#1})}

\newsavebox{\uuunit}
\sbox{\uuunit}
    {\setlength{\unitlength}{0.825em}
     \begin{picture}(0.6,0.7)
        \thinlines
        \put(0,0){\line(1,0){0.5}}
        \put(0.15,0){\line(0,1){0.7}}
        \put(0.35,0){\line(0,1){0.8}}
       \multiput(0.3,0.8)(-0.04,-0.02){10}{\rule{0.5pt}{0.5pt}}
     \end {picture}}
\newcommand {\unity}{\mathord{\!\usebox{\uuunit}}}
\begin{document}
\begin{titlepage}
\begin{flushright}
hep-th/0005219 \\
\end{flushright}
\vskip 2cm
\begin{center}
{{\Large \bf
Rings of short ${\cal N}=3$ superfields in three dimensions
and M-theory on $AdS_4\times \n010$. \hskip 0.2cm $^\dagger$
}}\\
\vfill
{\large M. Bill\'o$^1$, D. Fabbri$^1$,  P Fr\'e$^1$, P. Merlatti$^1$\\
and A. Zaffaroni$^2$} \\
\vfill
{ \sl $^1$ Dipartimento di Fisica Teorica, Universit\'a di Torino, via P.
Giuria 1,
I-10125 Torino, \\
 Istituto Nazionale di Fisica Nucleare (INFN) - Sezione di Torino,
Italy \\
$^2$ Istituto Nazionale di Fisica Nucleare (INFN) - Sezione di Milano,
Italy }
\end{center}
\vfill
\begin{abstract}
{In this paper we investigate three--dimensional superconformal
gauge theories with ${\cal N}=3$ supersymmetry.
Independently from specific models, we derive the shortening
conditions for unitary representations of the $\mathrm{Osp(3|4)}$
superalgebra and we express them in terms of differential
constraints on three dimensional ${\cal N}\!=\!3$ superfields.
We find a ring structure underlying these short representations,
which is just the direct generalization of the chiral ring structure
of ${\cal N}\!=\!2$ theories.
When the superconformal field theory is realized on the world--volume
of an M2--brane such superfield ring is the counterpart of the ring
defined by the algebraic geometry of the $8$--dimensional cone
transverse to the brane.
This and other arguments identify the ${\cal N}\!=\!3$ superconformal
field theory dual to M-theory compactified on AdS$_4\times \n010$.
It is an ${\cal N}=3$ gauge theory with $\mathrm{SU(N)}\times
\mathrm{SU(N)}$ gauge group coupled to a suitable set of
hypermultiplets, with an additional Chern Simons interaction.
The AdS/CFT correspondence can be directly verified using the
recently worked out Kaluza Klein spectrum of $\n010$ and we find
a perfect match.
We also note that besides the usual set of $\mathrm{BPS}$ conformal
operators dual to the lightest $\mathrm{KK}$ states, we find that
the composite operators corresponding to certain massive $\mathrm{KK}$
modes are organized into a massive spin $\ft 32$ ${\cal N}=3$
multiplet that might  be identified with the super-Higgs multiplet
of a spontaneously broken ${\cal N}\!=\!4$ theory.
We investigate this intriguing and inspiring feature in a separate paper.}
\end{abstract}
\vspace{2mm} \vfill \hrule width 3.cm
{\footnotesize
 $^ \dagger $ \hskip 0.1cm Supported by   EEC  under TMR contract
 ERBFMRX-CT96-0045}
\end{titlepage}
\section{Introduction}
There is  evidence that all M-theory or Type II string backgrounds
of the form $AdS_{p+2}\times X^{d-p-2}$ in d-dimensions, where
$X^{d-p-2}$ is an Einstein manifold, are dual to CFT's in $p+1$
dimensions, living on the world-volume of p-branes \cite{review}.
Many supergravity solutions associated with coset spaces  $X^{d-p-2}$
are known and have been studied in the eighties.
It is therefore interesting to identify the associated CFT's and compare
the KK spectrum with the spectrum of conformal operators.
The dual CFT is realized on the world-volume of p-branes living at the
singularities of $C(X^{d-p-2})$, the cone over $X^{d-p-2}$
\cite{witkleb,fig,morpless}.
Unfortunately, there is no general method for determining the world-volume
theory for branes in curved space-time, so one can only rely on geometrical
intuition.
Consider first the AdS$_5$ case.
There are only two supersymmetric five-dimensional coset spaces, $S^5$
and $T^{1,1}$. $S^5$ is at the origin of the original Maldacena conjecture.
The dual four-dimensional CFT dual to AdS$_5\times T^{1,1}$ have been
identified in \cite{witkleb}.
Many checks of this identification can be found in the literature
\cite{witkleb,gubser,gubserkleb,sergiotorino}.
In the AdS$_4$ case, there is a richer zoo of seven-dimensional coset spaces,
corresponding to supersymmetric backgrounds of M-theory \cite{castromwar}.
In \cite{3dcft}, we proposed candidate dual CFT's for the two ${\cal N}=2$
solutions, $Q^{1,1,1}$ \cite{q111} and $M^{1,1,1}$ \cite{m111}, using
intuition from toric geometry.
The KK spectrum and the properties of wrapped M5-branes associated to
baryons nicely fit with the CFT expectations.
A candidate dual for the ${\cal N}=2$ solution $V^{5,2}$, which does not
admit toric description, has been proposed in \cite{poliv52}.
The purpose of this paper, which is the natural continuation of \cite{3dcft},
is to discuss the  ${\cal N}=3$ solution AdS$_4\times \n010$.
\par
$\n010$ can be written as $\mathrm{SU(3)}/\mathrm{U(1)}$ \cite{castn010}.
It has isometry $\mathrm{SU(3)}\times \mathrm{SU(2)}$ and preserves
${\cal N}=3$ supersymmetry.
Using geometrical arguments similar to those in \cite{witkleb,3dcft},
one is led to consider an ${\cal N}=4$ theory $\mathrm{SU(N)}\times
\mathrm{SU(N)}$ with three hypermultiplets in the bi-fundamental
representation of the two gauge groups.
It was proposed in \cite{gukov} that the ${\cal N}=3$ CFT can be
just obtained by adding an ${\cal N}=3$ preserving Chern-Simons term.
We shall give evidence for this proposal by carefully comparing the
observables in the CFT and the excitations of the supergravity
background.
The complete KK spectrum of M-theory on AdS$_4\times N^{0,1,0}$
has been recently computed \cite{n010massspectrum,osp34}.
Both KK and conformal field theory composite operators fall in
representations of the superalgebra $\mathrm{Osp(3|4)}$ and can be
conveniently described in terms of three-dimensional ${\cal N}=3$
superfields.
In this paper, we first derive a general formalism for studying
${\cal N}=3$ superfields and the $\mathrm{Osp(3|4)}$ shortening
conditions and we then apply it to the comparison between KK states
and CFT composite operators.
We shall exhibit the CFT supermultiplets of composite operators
associated to all the short multiplets belonging to the KK
spectrum \footnote{One could also make an independent check of
the dimension of supersingletons in the CFT by looking at the
baryonic operators \cite{gubserkleb,3dcft}, which can be realized
as wrapped M5-branes. Since such a calculation would simply
be a repetition of known calculations that reveals no new feature
we skip such additional check, which should be straightforward.}.
Indeed the analysis of the ${\cal N}=3$ solution reveals that all
the general features which were common to the $T^{1,1}$, $Q^{1,1,1}$,
$M^{1,1,1}$ and $V^{5,2}$ compactifications \cite{gubser,sergiotorino,
3dcft,poliv52} still hold true also for $N^{0,1,0}$.
In particular, there are long multiplets with protected rational
dimensions.
We show that (in analogy with the other compactifications) many of
them can be identified with CFT multiplets obtained by tensoring
massless and short multiplets, as suggested in \cite{sergiotorino}.
We focus, in particular, on a very special long multiplet, which
contains the volume of the internal manifold as one of the scalar
components, and it is therefore {\it universal} for all compactification.
In ${\cal N}=2$ compactification, the volume multiplet is a long
vector multiplet.
In ${\cal N}=3$ it becomes a long gravitino multiplet.
In $\n010$, it has the right quantum numbers to be generated in a
superHiggs mechanism,  suggesting that the theory is a spontaneously
broken phase of an ${\cal N}=4 $ theory.
This intriguing phenomenon will be investigated in a forthcoming
publication \cite{noinext}.
\par
The plan of this work is as follows.
In section \ref{N=3supspace} we introduce the ${\cal N}\!=\!3$
superspace formalism and derive the shortening conditions of
the $\mathrm{Osp(3|4)}$ irreducible representations in terms of
differential constraints on primary conformal superfields.
In section \ref{N=3gaugetheory} we discuss the general structure of
${\cal N}=3$ three dimensional gauge theories using the component
formalism and emphasizing the role of the Chern Simons interaction.
In section \ref{theoryofN010} we identify the ${\cal N}=3$ gauge
theory whose conformal fixed point realizes the AdS/CFT correspondence
with the $\n010$ compactification of M-theory, while section
\ref{comparison} is devoted to test this correspondence.
Finally section \ref{suphigs} briefly discusses, from a CFT point of
view the long rational spin $\ft 32$ supermultiplet that suggests an
interpretation in terms of superHiggs mechanism and that will be the
focus of a forthcoming paper.
\section{Three dimensional ${\cal N}\!=\!3$ superspace}\label{N=3supspace}
In order to simplify the study of unitary irreducible representations
of the $\mathrm{Osp(3|4)}$ superconformal algebra (see eq. (A.2) of
\cite{susp}), we introduce a three dimensional ${\cal N}\!=\!3$
superspace formalism.
This allows us to identify the short representations as
particular constrained superfields.
To this effect we introduce six Grassmann coordinates,
$\theta^i_{\ \alpha}$, transforming as three Majorana bispinors
and as a triplet of the SO(3)$_R$ R-symmetry subalgebra:
\begin{eqnarray}
&[T^{ij}, T^{kl}] = -i\,(\delta^{jk}\,T^{il}-\delta^{ik}\,T^{jl}-
\delta^{jl}\,T^{ik}+\delta^{il}\,T^{jk})
\,, \nonumber \\
&T^{ij} = \theta^i_\alpha\frac{\partial}{\partial\theta_\alpha^j}-
\theta^j_\alpha\frac{\partial}{\partial\theta_\alpha^i}\,.
\end{eqnarray}
The other relevant generators have the following representation:
\begin{eqnarray}
&P_\mu=-i\partial_\mu\,,\nonumber\\
&q^{\alpha i}=\frac{\partial}{\partial\theta_\alpha^i}+
\frac{1}{2}\partial\!\!\!/^\alpha_{\ \beta}\theta^{\beta i}\,,\nonumber\\
&\{q^{\alpha i}, q^{\beta j} \} = - i\, \delta^{ij}\,
P\!\!\!\!/^{\ \alpha\beta}\,, \nonumber \\
&[T^{ij}, q^{\alpha k}] =
-i\, (\delta^{jk}\, q^{\alpha i} - \delta^{ik}\, q^{\alpha j} )\,.
\end{eqnarray}
We furthermore introduce the supercovariant derivatives:
\begin{eqnarray}
&{\cal D}^{\alpha i}=\frac{\partial}{\partial\theta_\alpha^i}-
\frac{1}{2}\partial\!\!\!/^\alpha_{\ \beta}\theta^{\beta i}\,,\nonumber\\
&\{{\cal D}^{\alpha i}, {\cal D}^{\beta j} \} = i\, \delta^{ij}\,
P\!\!\!\!/^{\ \alpha\beta}\,,\qquad\{q^{\alpha i}, {\cal D}^{\beta j}\}=0\,,
\end{eqnarray}
in terms of which the shortening conditions can be expressed.
\par
It is convenient to use the spherical irreducible basis of
R-symmetry representations rather than the cartesian one, so that
the Grassmann coordinates are renamed as in the following example:
\begin{equation}
\left\{\begin{array}{ccl}
\theta^+&=&-\ft{1}{\sqrt 2}(\theta^1+i\theta^2)\\
\theta^{\,0\,}&=&\theta^3\\
\theta^-&=&\ft{1}{\sqrt 2}(\theta^1-i\theta^2)
\end{array}\right.
\end{equation}
\par
An ${\cal N}\!=\!3$ superfield $\Theta=\Theta(x,\theta)$ is
a function of the bosonic $x^\mu$ and Grassmann coordinates
$\theta^i$, whose expansion in powers of $\theta^0$ gives us
the decomposition of the corresponding $\mathrm{Osp(3|4)}$
representation in ${\cal N}\!=\!2$ supermultiplets.
\par
We are mainly interested in conformal primary superfields
(see \cite{gunaydinminiczagerman2, macksalam}), defined by:
\begin{eqnarray}
\Theta(x, \theta) =\exp \left [\mbox{i}\, x_\mu P^\mu + \theta^i
q^i\right ]\Theta(0)\,,
\label{3Dsuperfield}
\end{eqnarray}
where $\Theta(0)$ is a primary field: $[s_\alpha^i,\Theta(0)]
=[K_\mu,\Theta(0)]=0$.
An irreducible representation of the superfield $\Theta(x, \theta)$
is characterized by the Cartan labels of its highest weight state
$\Theta(0)$, namely its scaling dimension, its SO(1,2) Lorentz and
SO(3)$_R$ R-symmetry quantum numbers.
We denote the R-symmetry isospin with the suffix $J$ and the
Lorentz character with a set of spinorial indices spanning the
SO(1,2) irreducible representation.
\subsection{Short $\mathrm{Osp(3|4)}$ representations as constrained
superfields}
In this section we analyze the differential constraints on the
superfields that force their components to transform into short
$BPS$ representations of the $\mathrm{Osp(3|4)}$ superalgebra.
\footnote{When this paper was nearly finished we learned of the
recent work by Ferrara and Sokatchev \cite{fersoka} that analyzes
the differential constraints to be imposed, in harmonic superspace,
on ${\cal N}=8$ superfields in order to describe short representations
of the algebra $\mathrm{Osp(8|4)}$. Quite likely our results for
$\mathrm{Osp(3|4)}$ can be described in that general formalism.
A comparison is postponed to future investigations}
As in the best known case of ${\cal N}=2$ superfields, we find the
existence of two kinds of constraints.
The first one is a first order differential constraint given by:
\begin{equation}
{\cal D}_{\alpha_1}\otimes_{h.w.}
\Theta^{J\ (\alpha_1\cdots\alpha_n)}(x,\theta)=0\,,
\label{diffconstraint}
\end{equation}
or
\begin{equation}
  {\cal D}_{\alpha}\otimes_{h.w.}\Theta^{J}(x,\theta)=0
\label{diffconszero}
\end{equation}
for scalar superfields (without Lorentz indices).
Here the tensor product refers to the SO(3)$_R$ isospin and ``h.w''
stands for highest weight.
This means that only the SO(3)$_R$ highest weight part of the tensor
product (\ref{diffconstraint}), between the isospin triplet ${\cal D}$
and the superfield $\Theta^J$, is put to zero.
\par
In complete analogy with the ${\cal N}\!=\!2$ case, we have another
kind of constraint.
It is a second order differential constraint, and it is allowed only
for scalar (from the Lorentz viewpoint) superfields (of any isospin):
\begin{equation}
{\cal D}_\alpha\otimes_{h.w.}{\cal D}^\alpha\otimes_{h.w.}\Theta^J=0\,.
\label{diffconstraint2}
\end{equation}
The superconformal covariance of equations (\ref{diffconstraint}) and
(\ref{diffconstraint2}) poses some constraints on the conformal
dimensions of the superfields.
\subsubsection{The ${\cal N}\!=\!3$ analogue of the chiral ring}
Let us now analyze in more details the constraint
(\ref{diffconszero}) for the lowest
Lorentz and isospin representations.
The most interesting case to analyze is that of a scalar (from
the Lorentz viewpoint) superfield.
In this case the Lorentz character allows the existence of a
{\bf ring structure} which generalizes the {\bf chiral ring} of
${\cal N}\!=\!2$ theories and seems to be a common feature shared
by all the three dimensional superconformal field theories.
The ring multiplicative operation is given by extracting  the
highest weight irreducible part from the ordinary (tensor) product
of two short superfields of isospin $J$ and $J'$ respectively:
\begin{equation}
\Theta^J\times\tilde\Theta^{J'}=\left(\Theta\otimes_{h.w.}
\tilde\Theta\right)^{J+J'}\,.
\end{equation}
Indeed one can show that:
\begin{equation}
\left.\begin{array}{c}
{\cal D}^{J\!=\!1}\otimes_{h.w.}\Theta^J=0\\
{\cal D}^{J\!=\!1}\otimes_{h.w.}\tilde\Theta^{J'}=0
\end{array}\right\}\Longrightarrow
{\cal D}^{J\!=\!1}\otimes_{h.w.}(\Theta\otimes_{h.w.}\tilde\Theta)
^{J+J'}=0\,.
\end{equation}
The simplest case of short scalar superfield, apart from the trivial
constant, is that of isospin $J\!=\!1/2$.
In this case  the shortening condition (\ref{diffconszero}) reads:
\begin{equation}
\left\{\begin{array}{c}
{\cal D}^+\Theta^+=0\,,\\
\sqrt{2}{\cal D}^0\Theta^++{\cal D}^+\Theta^-=0\,,\\
{\cal D}^-\Theta^++\sqrt{2}{\cal D}^0\Theta^-=0\,,\\
{\cal D}^-\Theta^-=0\,.
\end{array}\right.
\end{equation}
To make contact with ${\cal N}\!=\!2$ superspace formalism of
\cite{susp} it is useful to expand the most general form of
$\Theta^{J\!=\!1/2}$ in powers of $\theta^0$.
So we have:
\begin{equation}
\left(\begin{array}{c}
\Theta^+\\
\Theta^-
\end{array}\right)=\left(\begin{array}{c}
\Phi_s\\
\Psi_s^\dagger\end{array}\right)
-\ft{1}{\sqrt 2}
\left(\begin{array}{c}
{\cal D}^+\Psi_s^\dagger\\
{\cal D}^-\Phi_s
\end{array}\right)
\theta^0
\,,\label{J=1/2chiral}
\end{equation}
where $\Phi_s$ and $\Psi_s$ are two ${\cal N}\!=\!2$
{\bf supersingletons}, namely they are two functions of $x^\mu$
and $\theta^\pm$ fulfilling the constraints ${\cal D}^+\Phi_s=
{\cal D}^-{\cal D}^-\Phi_s=0$.
Hence we see that the direct generalization of the ${\cal N}\!=\!2$
supersingleton is the ${\cal N}\!=\!3$ short scalar superfield of
minimum isospin.
Let us now look at the case $J\!=\!1$, whose most general form is:
\begin{equation}
\left(\begin{array}{c}
\Theta^+\\
\Theta^{\,0\,}\\
\Theta^-
\end{array}\right)=\left(\begin{array}{c}
\Phi\\
\Sigma\\
\Psi^\dagger
\end{array}\right)-\left(\begin{array}{c}
{\cal D}^+\Sigma\\
\ft{1}{2}({\cal D}^-\Phi+{\cal D}^+\Psi^\dagger)\\
{\cal D}^-\Sigma
\end{array}\right)\theta^0-\ft{1}{8}\left(\begin{array}{c}
{\cal D}^+{\cal D}^+\Psi^\dagger\\
2{\cal D}^+{\cal D}^-\Sigma\\
{\cal D}^-{\cal D}^-\Phi
\end{array}\right)(\theta^0\theta^0)\,,\label{J=1chiral}
\end{equation}
where $\Phi$ and $\Psi$ are ${\cal N}\!=\!2$ chiral superfields
and $\Sigma$ is a linear superfield (${\cal D}^+{\cal D}^+\Sigma=
{\cal D}^-{\cal D}^-\Sigma=0$), which is a conserved (massless)
vector current.
Hence the superfield (\ref{J=1chiral}) represents the direct
generalization of the ${\cal N}\!=\!2$ massless vector.
\par
The ${\cal N}\!=\!3$ short scalar superfields of higher isospin can
be obtained by multiplying smaller ones following the ring operation,
i.e. by tensoring and taking the maximum isospin irreducible part.
It is interesting to analyze the ${\cal N}\!=\!2$ field content,
i.e. the single independent $\theta^0$ components of such superfields.
This gives an analytical version of the algebraic ${\cal N}\!=\!3\to
{\cal N}=2$ decomposition of the short multiplets (see tables
\eqn{N=3massless} and \eqn{N=3short}).
The first thing to note is that the shortening constraint
\eqn{diffconstraint} implies that the only independent components are
the $\theta^0\!=\!0$ restrictions of the ${\cal N}\!=\!3$ superfields.
In the case of integer isospin we always obtain the same pattern:
\begin{equation}
\left(\begin{array}{c}
\Phi^1\Phi^2\ldots\Phi^k\\
\Sigma^1\Phi^2\ldots\Phi^k+\ldots+
\Phi^1\ldots\Phi^{(k-1)}\Sigma^k\\
\ldots\\
\Sigma^1\Psi^{2\dagger}\ldots\Psi^{k\dagger}+\ldots+
\Psi^{1\dagger}\ldots\Psi^{(k-1)\dagger}\Sigma^k\\
\Psi^{1\dagger}\Psi^{2\dagger}\ldots\Psi^{k\dagger}
\end{array}\right)^{J\!=\!k}_{\theta^0=0}\ \
\left.\begin{array}{cc}
\leftarrow&{\rm chiral}\\
\leftarrow&{\rm short\ vector}\\
\leftarrow&2k\!-\!1\ {\rm long\ vectors}\\
\leftarrow&{\rm short\ vector}\\
\leftarrow&{\rm chiral}\\
\end{array}\right.
\end{equation}
The half-integer isospin chiral superfields have a completely
analogous structure.
The only difference is that each field contains an odd number of
${\cal N}\!=\!2$ supersingletons.
Thus the corresponding states are not observed in the Kaluza Klein
spectrum of supergravity compactifications.
\subsubsection{The ${\cal N}\!=\!3$ short gravitinos}
Let us now analyze the second order constraint (\ref{diffconstraint2}),
which yields the ${\cal N}\!=\!3$ short gravitinos.
The lowest isospin case ($J=0$) corresponds to the massless gravitino
superfield:
\begin{equation}
\Theta=\Sigma+G_\alpha\theta^{0\,\alpha}+\ft{1}{4}({\cal D}^+
{\cal D}^-)\Sigma(\theta^0\theta^0)\,,
\end{equation}
where $G^\alpha$ is an ${\cal N}\!=\!2$ massless gravitino
(${\cal D}^\pm_\alpha G^\alpha=0$) and $\Sigma$ a linear superfield,
namely a massless vector.
\par
Analogously we can derive the form of the most general
$J\!=\!1$ short spinor superfield:
\begin{eqnarray}
\left(\begin{array}{c}
\Theta^+\\
\Theta^{\,0\,}\\
\Theta^-
\end{array}\right)=
\left(\begin{array}{c}
\Sigma^+\\
\Sigma^{\,0\,}\\
\Sigma^-
\end{array}\right)+
\left(\begin{array}{c}
G^+\theta^0\\
G^{\,0\,}\theta^0\\
G^-\theta^0
\end{array}\right)+{\rm derivative\ terms}\,,
\label{J=1shgravino}
\end{eqnarray}
which has six ${\cal N}\!=\!2$ independent components:
\begin{itemize}
\item two short  gravitinos
(${\cal D}^+G^+={\cal D}^-G^-=0$)\,;
\item one long  gravitino, $G^0$\,;
\item two short vectors: $({\cal D}^+{\cal D}^+)\Sigma^+
=({\cal D}^-{\cal D}^-)\Sigma^-=0$\,;
\item one long vector, $\Sigma^0$\,.
\end{itemize}
This ${\cal N}\!=\!2$ superfield content perfectly fits the
algebraic decomposition of table \eqn{N=3short}.
Short gravitinos of higher isospin can be obtained by composing
the $J\!=\!0$ short gravitino with chiral superfields of any $J$.
Obviously, even in this case, half-integer isospin gravitinos are
not observed in the Kaluza Klein spectra, due to the presence of
an odd number of supersingletons.
\subsubsection{The ${\cal N}\!=\!3$ short gravitons}
The ${\cal N}\!=\!3$ short graviton multiplets are realized by spinor
superfields fulfilling the first order constraint (\ref{diffconstraint}).
Again, the massless case corresponds to the lowest ($J=0$) isospin
superfield:
\begin{equation}
\Theta^\alpha=G^\alpha+T^{(\alpha\beta)}\theta^0_{\,\beta}-\ft{1}{4}
\partial\!\!\!/^\alpha_{\ \beta}G^\beta(\theta^0\theta^0)\,,
\label{J=0shgraviton}
\end{equation}
where $G^\alpha$ is an ${\cal N}\!=\!2$ massless gravitino
(${\cal D}^\pm_\alpha G^\alpha=0$) and $T^{(\alpha\beta)}$
is a massless graviton (${\cal D}^\pm_\alpha T^{(\alpha\beta)}=0$).
\par
In an analogous way we can derive the form of the most general
$J\!=\!1$ short spinor superfield:
\begin{eqnarray}
\left(\begin{array}{c}
\Theta^{+\,\alpha}\\
\Theta^{\,0\,\,\alpha}\\
\Theta^{-\,\alpha}
\end{array}\right)=
\left(\begin{array}{c}
G^{+\,\alpha}\\
G^{\,0\,\,\alpha}\\
G^{-\,\alpha}
\end{array}\right)+
\left(\begin{array}{c}
T^{+(\alpha\beta)}\\
T^{\,0\,(\alpha\beta)}\\
T^{-(\alpha\beta)}
\end{array}\right)\theta^0_{\,\beta}+{\rm derivative\ terms}\,,
\end{eqnarray}
which has six ${\cal N}\!=\!2$ independent components
(see table \eqn{N=3short}):
\begin{itemize}
\item two short gravitons
(${\cal D}^+_\alpha T^{+(\alpha\beta)}=
{\cal D}^-_\alpha T^{-(\alpha\beta)}=0$)\,;
\item one long graviton, $T^{\,0\,(\alpha\beta)}$\,;
\item two short gravitinos
(${\cal D}^+G^+={\cal D}^-G^-=0$)\,;
\item one long gravitino, $G^0$\,.
\end{itemize}
Short gravitons of higher isospin can be obtained by composing
the $J=0$ massless graviton with chiral superfields of any $J$.
Again, half-integer isospin gravitons, containing an odd number of
supersingletons, are not observed in the Kaluza Klein spectra.
\section{${\cal N}=3$ gauge theory in three dimensions}\label{N=3gaugetheory}
In this section we discuss the structure of a three dimensional gauge
theory with ${\cal N}=3$ supersymmetry. In paper \cite{susp} we
have already given the general form of an ${\cal N}=2$
three--dimensional gauge theory and the ${\cal N}=3$ case is just
a particular case in that class since a theory with ${\cal N}=3$
SUSY, must {\it a fortiori} be an ${\cal N}=2$ theory.
In \cite{susp} we have also considered, within the ${\cal N}=2$
class, the case of ${\cal N}=4$ theories. These are obtained
through dimensional reduction of an ${\cal N}_4=2$ theory in
four--dimensions.
Indeed since each $D=4$ Majorana spinor splits, under dimensional
reduction on a circle $\mathbb{S}^1$, into two $D=3$ Majorana
spinors, the number of three--dimensional supercharges is just twice
the number of $D=4$ supercharges:
\begin{equation}
  {\cal N}_3 = 2 \, \times \, {\cal N}_4
\label{n34}
\end{equation}
The ${\cal N}_3=3$ case corresponds to an intermediate situation.
It is an ${\cal N}_3=2$ theory with the field content of an
${\cal N}_3=4$ one, but with additional ${\cal N}_3=2$ interactions
that respect  three out of the four supercharges obtained through
dimensional reduction.
Using an ${\cal N}=2$ superfield formalism and the notion of twisted
chiral multiplets it was shown in \cite{kapustin} that for abelian
gauge theories these additional ${\cal N}_3=3$ interactions are
\begin{enumerate}
  \item A Chern Simons term, with coefficient $\alpha$
  \item A mass-term  with coefficient $\mu=\alpha$ for the
  chiral field $Y^I$ in the adjoint of the color gauge group.
  By this latter we denote the complex field belonging,
  in four dimensions, to the ${\cal N}_4=2$ gauge
  vector multiplet.
\end{enumerate}
In this section we want to retrieve the same result in the
component formalism which is better suited to discus the relation
between the world--volume gauge theory and the geometry of the
transverse cone $\mathcal{C}(X^7)$.
Then we dismiss superfields and turning to components we discuss
the general form of a non abelian $ {\cal N}=3$ gauge theory in
three dimensions.
\subsection{The field content and the interactions}
Our strategy is that of writing the ${\cal N}=3$ gauge theory
as aspecial case of an ${\cal N}=2$ theory, whose general form
was derived in \cite{susp}. For this latter the field content
is given by:
\begin{equation}
   \begin{array}{|c|c|c|c|}
   \hline
     \mbox{multipl. type $\, /\,SO(1,2)$ spin}  &  1 & \ft 1 2 & 0 \\
     \hline
     \hline
     \null & \null &\null & \null \\
     \mbox{vector multipl.} &  \underbrace{A^I_\mu}_{\mbox{gauge field}} &
     \underbrace{\left( \lambda^{+I},\lambda^{-I}
     \right)}_{\mbox{gauginos}}  &\underbrace{ M^I}_{\mbox{real scalar}} \\
     \hline
     \null &  \null &\null & \null \\
     \mbox{chiral multip.} &  \null &\underbrace{\left( \chi^{+i},\chi^{-i^*}
     \right)}_{\mbox{chiralinos}}  & \underbrace{ z^i,\ \bar z^{i^*}}_
     {\mbox{complex scalars}}
     \\
     \hline
   \end{array}
\label{fieldcont}
\end{equation}
and without Fayet Iliopoulos terms, which do not exist in
non abelian gauge theories with no U(1) factors,
the complete Lagrangian has the following form:
\begin{equation}\label{N=2stLag}
{\cal L}^{{\cal N}=2}={\cal L}^{kinetic}
+{\cal L}^{fermion~mass}+{\cal L}^{potential}\,,
\end{equation}
where
\begin{eqnarray}\label{N=2chiralst}
\mathcal{L}^{kinetic}&=&\left\{
\eta_{ij^*}\nabla_\mu z^i\nabla^\mu\overline z^{j^*}
-\ft{1}{2}\eta_{ij^*}\left(\chi^{-j^*}{\nabla\!\!\!\!/}\chi^{+i}
+\chi^{+i}{\nabla\!\!\!\!/}\chi^{-j^*}\right)
\right.\nonumber\\
&&-g_{IJ}F^I_{\mu\nu}F^{J\,\mu\nu} - \alpha \left( g_{IJ}
F^I_{\mu\nu}A^J_\rho + f_{IJK} A^I_\mu A^J_\nu A^K_\rho \right)  \,
\epsilon^{\mu\nu\rho}\nonumber\\
&&+\left.\ft{1}{2}g_{IJ}\nabla_\mu M^I\nabla^\mu M^J
-\ft{1}{4}g_{IJ}\left(\l^{-I}{\nabla\!\!\!\!/}\l^{+J}
+\l^{+I}{\nabla\!\!\!\!/}\l^{-J}\right)\right\}d^3x\\
&&\nonumber\\
\mathcal{L}^{fermion~mass}&=&
\left\{\ft{i}{2}\left(\chi^{+i}\partial_i\partial_jW(z)\chi^{+j}
-\chi^{-i^*}\partial_{i^*}\partial_{j^*}\overline W(\overline z)
\chi^{-j^*}\right)\right.\nonumber\\
&&-\ft{i}{2}f_{IJK}M^I\l^{-J}\l^{+K}
-i\chi^{-j^*}M^I(T_I)_{ij^*}\chi^{+i}\nonumber\\
&&-\left(\chi^{-i^*}\l^{+I}(T_I)_{i^*j}z^j
-\chi^{+i}\l^{-I}(T_I)_{ij^*}\overline z^{j^*}\right)\nonumber\\
&&\left.-\ft{1}{2}\alpha g_{IJ}\l^{-I}\l^{+J}\right\}d^3x\\
\mathcal{L}^{potential}&=&-U(z,\overline z)d^3x\,,
\end{eqnarray}
the scalar potential admitting the following general expression
\begin{eqnarray}
U(z,\overline z,M)&=&
\partial_i W(z)\eta^{ij^*}\partial_{j^*}\overline W(\overline z)\nonumber\\
&&+\ft{1}{2}g^{IJ}\left(\overline z^{i^*}(T_I)_{i^*j}z^j\right)
\left(\overline z^{k^*}(T_J)_{k^*l}z^l\right)\nonumber\\
&&+\overline z^{i^*}M^I(T_I)_{i^*j}\eta^{jk^*}M^J(T_J)_{k^*l}
z^l\nonumber\\
&&2\alpha^2g_{IJ}M^IM^J
  -2\alpha M^I\left(\overline z^{i^*}(T_I)_{i^*j}z^j\right)
  \label{scalapote}
\end{eqnarray}
and the {\it superpotential} $W(z)$ being an arbitrary holomorphic
function of the chiral scalars $z^i$. Our index notations
and conventions are given in the appendices.
\par
The ${\cal N}=3$ case is obtained when the  following conditions
are fulfilled:
\begin{itemize}
  \item The spectrum of chiral multiplets and their representation assignments
   under the gauge and flavor groups are as follows
  \begin{equation}
  z^i=\left\{\begin{array}{l lcl}
    \sqrt{2} \,Y^I & \mbox{\bf adj} \, \left[\mathcal{G}_{gauge}\right]
    & \times & \mbox{\bf id} \,
    \left[\mathcal{G}_{flavor}\right]\\
    g\,u^a & \mathbf{R}_g \, \left[\mathcal{G}_{gauge} \right] &
    \times & \mathbf{R}_f  \,\left[ \mathcal{G}_{flavor}\right]\\
    g\,v_a & \overline{{\mathbf{R}}}^{-1}_g \,
    \left[\mathcal{G}_{gauge} \right] & \times &
    \overline{{\mathbf{R}}}_f^{-1} \,\left[ \mathcal{G}_{flavor}\right]
  \end{array}\right. \, \Rightarrow \,
  \eta^{ik^*}T^I_{k^*j}=\left\{\begin{array}{l}
    i \, f^I_{\,JK}\\
    (T^I)_{a}^{\phantom{a}b}\\
    -(\overline T^I)_{a}^{\phantom{a}b}
  \end{array}\right.
  \label{istcond}
  \end{equation}
  $\mathbf{R}_g$, and $\mathbf{R}_f$ being two complex representations
  of $G_{gauge}$ and $G_{flavor}$, respectively.
  \item The superpotential $W(z)$ has the following form:
  \begin{equation}
  W(Y,u,v)=g_{IJ}\,\left(2\,g\,Y^I\,v_a\,T^{J\vert a}_{\quad\ b}\,
  u^b\,+\,2\,\alpha\,\,Y^I\,Y^J\right)
  \label{suppotn3}
  \end{equation}
\end{itemize}
The reason why these two choices make the theory ${\cal N}_3=3$
invariant is simple: the first choice corresponds to assuming the
field content of an ${\cal N}_3=4$ theory which is necessary since
${\cal N}_3=3$ and ${\cal N}_3=4$ supermultiplets are identical.
The second choice introduces an interaction that preserves ${\cal N}_3=3$
supersymmetry but breaks (when $\alpha \ne 0$) ${\cal N}_3=4$
supersymmetry. We can appreciate the last statement if we rewrite the
Lagrangian in such a way that its invariance against the
$\mathrm{so(3)_R}$ R-symmetry becomes manifest. To this effect we begin
by recalling that viewed from an ${\cal N}_3=3$ or
${\cal N}_3=4$ vantage point the chiral fields $u^a,v_a$ are the bosonic
elements of a hypermultiplet and can be organized into a quaternion,
according to the rule:
\begin{equation}
  Q^a = \left( \begin{array}{cc}
    u^a & i\,\overline{{ v}}^a \\
    i\,v_a & \overline{{ u}}_a \
  \end{array}\right) \, \equiv \, q^{a|0}\unity+iq^{A|x}\sigma_x
\label{quaternio}
\end{equation}
In this way the transformation of the hypermultiplet $u^a, v_a$ under
gauge or flavor generators can be rewritten as follows:
\begin{eqnarray}
\delta^I{\bf Q}&=&i\hat T^I{\bf Q}\nonumber\\
&&\nonumber\\
\delta^I\left(\begin{array}{cc}
u^a&i\overline v^{a}\\
iv_a&\overline u_{a}
\end{array}\right)&=&
i\left(\begin{array}{cc}
T^{I\vert a}_{\phantom{I\vert a}b}&\\
&-\overline T^{I\phantom{a}b}_{\phantom{I}a}
\end{array}\right)\left(\begin{array}{cc}
u^b&i\overline v^{b}\\
iv_b&\overline u_{b}
\end{array}\right)
\end{eqnarray}
where the $T^{I\vert a}_{\phantom{I\vert a}b}$ realize a representation
of ${\cal G}$ in terms of $n\times n$ hermitian matrices.
We define $\overline{{ T}}^{I\phantom{a}b}_{\phantom{I}a}
\equiv\left(T^{I\vert a}_{\phantom{I\vert a}b}\right)^*$.
\par
Under the SO(3)$_R$ R--symmetry the hypermultiplets transform as an SU(2)
doublet, in the sense that for each $\mathcal{U}\in$ SU(2)$_R \sim $SO(3)$_R$
the quaternion varies as follows:
\begin{equation}
  \delta_R \, Q^a = Q^a \, \mathcal{U}
\label{delquate}
\end{equation}
On the other hand the auxiliary fields that appear in the gaugino's
supersymmetry transformation rules vary, under R--symmetry in the
triplet representation of SO(3).
Their on--shell values constitute the so called triholomorphic momentum
map. This is a unimodular quaternion bilinear constructed by means of
the gauge group generators. Explicitly one sets:
\begin{equation}
  {\cal P}^I=\ft 1 2 \, i \, \left( {\bar {\bf Q}} \, {\hat T}^I \,{\bf Q}
  \right) = \left( \begin{array}{cc}
    {\cal P}^I_3 & {\cal P}^I_+ \\
    {\cal P}^I_-  & -{\cal P}^I_3 \
  \end{array}\right)
\label{triholo}
\end{equation}
where:
\begin{eqnarray}
{\cal P}^I_3  & = & -\left( {\bar u}_a \,
T^{I\vert a}_{\phantom {I\vert a}b}\, u^b- {\bar v}^a \,
{\bar T}^{I\vert\phantom{a} b}_{\phantom{I\vert}a}\,
v_b \right)  \nonumber\\
{\cal P}^I_-  & = &2\delta_{ac}{\bar v}^c\, T^{I\vert
a}_{\phantom{I\vert a}b} \, u^b =
2 {  v}_a \, T^{I\vert
a}_{\phantom{I\vert a}b} \, u^b  \nonumber\\
{\cal P}^I_+  & = & -\left({\cal P}^I_-\right)^* =
-2 {\bar  v}^a \, {\bar T}^{I\vert\phantom{a} b}_{\phantom{I\vert}a}
\, {\bar u}_b
\label{P3+-}
\end{eqnarray}
The first form of ${\cal P}^I_-$ explicitly exhibits the SU(2)
covariance in the sense that $(u^a,\bar{v}^a)$ is a doublet.
The second expression will be interpreted later.
Out of the triholomorphic momentum map we extract the three components
of a real SO(3)$_R$ trivector.
Explicitly we set:
\begin{eqnarray}
  {\cal P}^I_\ell &\equiv& \left\{
  {\cal P}^I_1,\,{\cal P}^I_2,\,
  {\cal P}^I_3 \right \} \nonumber\\
  {\cal P}^I_-&=&-i \left({\cal P}^I_1 +{\cal P}^I_2 \right)\nonumber\\
  {\cal P}^I_+&=&-i \left({\cal P}^I_1 -{\cal P}^I_2 \right)
\label{ppp}
\end{eqnarray}
There is another SO(3)$_R$  real trivector in the theory which is composed by
the complex scalar field $Y^I$ in the adjoint representation of the gauge group
together with the real scalar $M^I$ belonging to the ${\cal N}=2$
vector multiplet. Explicitly we set:
\begin{equation}
  \phi_\ell^I= \left( \begin{array}{c}
    -\mbox{Im}Y^I \\
    \mbox{Re} Y^I \\
    \ft 12 \, M^I \
  \end{array}\right)
\label{trivect}
\end{equation}
Inserting eq.(\ref{suppotn3}) into the general ${\cal N}_3=2$ formula
(\ref{scalapote}) and using
the notations of eq.s(\ref{ppp},\ref{trivect}) we can rewrite the final form
of the ${\cal N}=3$ scalar
potential in a way that exhibits manifest invariance under  $\mathrm{so(3)}$
R-symmetry and is a sum of squares:
\begin{eqnarray}
U & = & g_{IJ} \, \delta^{\ell m} \,
\left [ 2 \sqrt{2} \, \alpha \, \phi^I_\ell + \ft {1}{\sqrt{2}} \, g \,
{\cal P}^I_\ell+\mathcal{Q}^I_\ell\right]
\,
\left [ 2 \sqrt{2} \, \alpha \, \phi^J_m + \ft {1}{\sqrt{2}} \, g \, {\cal P}^J_m
+\mathcal{Q}^J_m\right]
\nonumber\\
&&+4\, g^2 \, g_{IJ} \, \delta^{\ell m} \, \phi^I_\ell \, \phi^J_m \, \left[
{\bar u}_a \, (T^I \, T^J )^{a}_{\phantom{a}b} \, u^b +
 {\bar v}^a ({\bar T}^I {\bar T}^J )_{a}^{\phantom{a}b} v_b
\right]
\label{finopote}
\end{eqnarray}
where:
\begin{equation}
  \mathcal{Q}^I_\ell=   \sqrt{2}\, \epsilon_{\ell m n} \, \phi^P_m
  \,\phi^Q_n \, f^I_{\phantom{I}PQ}
\label{giorgio}
\end{equation}
The classical vacua of the ${\cal N}=3$ theory are immediately
determined from eq.(\ref{finopote}). One has:
\begin{eqnarray}
  \phi^I_\ell &=&0   \label{liftcoulo}  \\
   {\cal P}^J_\ell (u,v) &=& 0 \label{hypkal}
\end{eqnarray}
Eq. (\ref{liftcoulo}) lifts the Coulomb branch of the
theory setting to zero the vev.s of the scalar fields in the
adjoint representation of the gauge group. Eq.
(\ref{hypkal}), instead identifies the manifold of classical vacua with the
HyperK\"ahler quotient of the flat HyperK\"ahler manifold spanned by
the hypermultiplets $u^a,\bar{v}^a$ with respect to the
triholomorphic action of the gauge group.
The locus defined by
(\ref{hypkal}) is the zero level set of the triholomorphic momentum
map and it has to be further modded out by the action of U(1).
When the generator $ T^{I\vert a}_b={\rm i} \delta^a_b$ is a
U(1)--generator, eq.s (\ref{P3+-}) just reproduce the definition
of the flag variety ${\mathbb F}(1,2;3) \simeq$ SU(3)/U(1)$\times$U(1)
which is the base manifold of $N^{0,1,0}$ seen as a circle bundle (see
\cite{3dcft}). This is what we explain in more details in the next section.
\section{The ${\cal N}=3$ gauge theory corresponding to the
$\n010$ compactification}\label{theoryofN010}
Having clarified the structure of a generic ${\cal N}=3$ gauge
theory let us consider the specific one associated with the
$\n010$ seven--manifold.
As explained in \cite{3dcft} (see eq.(B.1) of that paper)
the manifold $\n010$ is the circle bundle inside $\mathcal{O}(1,1)$
over the flag manifold $\mathbb{F}(1,2;3)$. In other words we have
\begin{equation}
  \n010 \, \stackrel{\pi}{\longrightarrow}\mathbb{F}(1,2;3)
\label{fibrsuflag}
\end{equation}
where, by definition,
\begin{equation}
  \mathbb{F}(1,2;3) \equiv \frac{\mathrm{SU(3)}}{H_1 \times H_2}
\label{flagga}
\end{equation}
is  the homogeneous space obtained by modding $\mathrm{SU(3)}$
with respect to its maximal torus:
\begin{equation}
  H_1 = \exp \left[i \theta_1 \left( \begin{array}{ccc}
    1 & 0 & 0 \\
    0 & -1 & 0 \\
    0 & 0 & 0 \
  \end{array}\right) \right] \quad ; \quad
   H_2 = \exp \left[i \theta_2 \left(\begin{array}{ccc}
     1 & 0 & 0 \\
     0 & 1 & 0 \\
     0 & 0 & -2 \\
   \end{array} \right) \right]
\label{maxtor}
\end{equation}
Furthermore as also explained in \cite{3dcft} (see
eq.(B.2)), the base manifold $\mathbb{F}(1,2;3)$ can be algebraically
described as the following quadric
\begin{equation}
  \sum_{i=1}^{3} \, u^i \, v_i = 0
\label{vanlocus}
\end{equation}
in $\mathbb{P}^2 \times \mathbb{P}^{2*}$, where $u^i$ and $v_i$ are
the homogeneous coordinates of
$\mathbb{P}^2$ and $\mathbb{P}^{2*}$, respectively.
\par
Hence a complete description of the metric cone $\mathcal{C}\left(
\n010\right) $ can be given by writing the following equations
in $ \mathbb{C}^3 \times \mathbb{C}^{3*}$:
\begin{equation}
 \mathcal{C}\left(
 \n010\right)= \left \{ \begin{array}{rcll}
    |u^i|^2-|v_i|^2 & = & 0 & \mbox{fixes equal the radii of $\mathbb{P}^2$
    and $\mathbb{P}^{2*}$}  \\
    2 \, u^i \, v_i & = & 0 & \mbox{cuts out the quadric locus} \\
    \left( u^i \, e^{i\theta} , v_i \, e^{-i\theta}\right)
    &\simeq &\left( u^i,v_i\right) &\mbox{identifies points of $\mathrm{U(1)}$
    orbits}\
  \end{array}\right.
\label{metrcon}
\end{equation}
Eq.s (\ref{metrcon}) can be easily interpreted as the statement that
the cone ${\cal C}\left(\n010\right)$ is the HyperK\"ahler quotient of a
flat three-dimensional quaternionic space with respect to the
triholomorphic action of a $\mathrm{U(1)}$ group.
Indeed the first two equations in (\ref{metrcon}) can be rewritten as the
vanishing of the triholomorphic momentum map of a $\mathrm{U(1)}$ group.
It suffices to identify:
\begin{eqnarray}
{\cal P}_3 & = & -\left( |u^i|^2 -|v_i|^2 \right)  \nonumber\\
{\cal P}_- & =& 2 v_i u^i
\label{agniu1}
\end{eqnarray}
Comparing with eq.s (\ref{liftcoulo}) we see that the cone $\mathcal{C}
(\n010)$ can be correctly interpreted as the space of classical vacua
in an abelian ${\cal N}=3$ gauge theory with $3$ hypermultiplets in
the fundamental representation of a flavor group $\mathrm{SU(3)}$.
\par
This suggests that the ${\cal N}=3$ non--abelian gauge theory whose
infrared conformal point is dual to supergravity compactified on
AdS$_4 \times \n010$ has the following structure:
\begin{equation}
 \begin{array}{crcl}
   \mbox{gauge group} & \mathcal{ G}_{gauge} & = & \mathrm{SU(N)}_1 \times
   \mathrm{SU(N)}_2 \\
   \null & \null & \null & \null \\
   \mbox{flavor group} & \mathcal{G}_{flavor} & = & \mathrm{SU(3)}  \\
   \null & \null & \null & \null \\
   \mbox{color representations of hypermultiplets} & \left[ \begin{array}{c}
  u  \\
  v
\end{array} \right] & \Rightarrow & \left[ \begin{array}{c}
  \left({\bf N}_1,{\bf \bar N}_2\right)  \\
  \left({\bf \bar N}_1, {\bf N}_2\right)
\end{array} \right]  \\
\null & \null & \null & \null \\
\mbox{flavor representations of hypermultiplets} & \left[ \begin{array}{c}
  u  \\
  v
\end{array} \right] & \Rightarrow & \left[ \begin{array}{c}
  \left({\bf 3}, {\bf \bar 3}\right)  \\
 \left({\bf \bar 3}, {\bf 3} \right)
\end{array} \right]  \\
 \end{array}
\label{assegnati}
\end{equation}
More explicitly and using an ${\cal N}=2$ notation we can say that
the field content of our theory is given by the following chiral
fields, that are all written as $N \times N$ matrices:
\begin{equation}
  \begin{array}{cccc}
    Y_1 & = & \left(Y_1\right)^{\Lambda_1}_{\phantom{\Lambda_1}\Sigma_1 } &
    \mbox{adjoint of $\mathrm{SU(N)}_1$} \\
   Y_2 & = & \left(Y_2\right)^{\Lambda_2}_{\phantom{\Lambda_2}\Sigma_2}  &
    \mbox{adjoint of $\mathrm{SU(N)}_2$ }\\
     u^i & = & \left(u^i\right)^{\Lambda_1}_{\phantom{\Lambda_1}\Sigma_2 } &
    \mbox{in the $({\bf 3},{\bf N}_1,{\bf \bar N}_2)$} \\
    v_i & = & \left(v_i\right)^{\phantom{\Sigma_1}\Lambda_2}_{\Sigma_1}  &
    \mbox{in the $({\bf 3},{\bf \bar N}_1,{\bf N}_2)$} \\
  \end{array}
\label{matricicole}
\end{equation}
and the superpotential can be written as follows:
\begin{equation}
  W = 2 \, \left [ g_1 \mbox{Tr} \left( Y_1\, u^i \, v_i \right) + g_2
  \mbox{Tr}\left( Y_2\, v_i \, u_i \right) +  \alpha_1 \mbox{Tr}\left(
  Y_1 \, Y_1 \right ) + \alpha_2 \, \mbox{Tr} \left( Y_2 \, Y_2 \right) \right]
\label{suppotmat}
\end{equation}
where $g_{1,2},\alpha_{1,2}$ are the gauge coupling constants and
Chern Simons coefficients associated with
the $\mathrm{SU(N)}_{1,2}$ simple
gauge groups, respectively.
Setting:
\begin{eqnarray}
g_1 & = & g_2 = g\nonumber\\
\alpha_1 & = & \pm \alpha_2 = \alpha
\label{galpha}
\end{eqnarray}
and integrating out the two fields $Y_{1,2}$ that have received a
mass by the Chern Simons mechanism we obtain the effective quartic
superpotential:
\begin{equation}
  W^{eff}= -\ft 1 2 \, \ft {g^2}{\alpha} \left[
  \mbox{Tr} \left( v_i \, u^i \, v_j \, u^j  \right) \pm
  \mbox{Tr} \left( u^i \, v_i \, u^j \, u_j\right)   \right]
\label{effepot}
\end{equation}
The vanishing relation one obtains from the above superpotential are
the following ones:
\begin{equation}
  u^i \, v_j \, u^j = \pm u^j \, v_j \,  u^i \quad ; \quad
   v_i \, u^j \, v_j = \pm v_j \, u^j \,  v_i
\label{vanrel}
\end{equation}
Consider now the chiral conformal superfields one can write in this
theory:
\begin{equation}
  \Phi^{i_1\, i_2 \, \dots \, i_k}_{j_1 \, j_2 \, \dots \, j_k}
  \equiv
  \mbox{Tr} \left( u^{(i_1} \, v_{(j_1} \, u^{i_2} \, v_{j_2} \, \dots
  \, u^{i)_k} \, v_{j_k)} \right)
\label{chiralop}
\end{equation}
where the round brackets denote symmetrization on the indices.
The above operators have $k$ indices in the fundamental
representation of $\mathrm{SU(3)}$ and $k$ indices in the antifundamental one,
but they are not yet assigned to the irreducible representation:
\begin{equation}
  M_1=M_2 = k
\label{irredu}
\end{equation}
as it is predicted both by general geometric arguments and by the
explicit evaluation of the Kaluza Klein spectrum of hypermultiplets
\cite{osp34}. To be irreducible the operators (\ref{chiralop}) have
to be traceless. This is what is implied by the vanishing relation
(\ref{vanrel}) if we choose the minus sign in eq.(\ref{galpha}).
\par
Notice that for $N^{0,1,0}$ the form of the superpotential, which is
dictated by the Chern-Simons term, is strongly reminiscent of the
superpotential considered in \cite{witkleb}.
The CFT theory associated with  $N^{0,1,0}$ has indeed many analogies
with the simpler cousin $T^{1,1}$.
There is however also a crucial difference.
We recognize a general phenomenon that we already discussed in the
$M^{1,1,1}$ and $Q^{1,1,1}$ compactifications \cite{3dcft}.
The moduli space of vacua of the abelian theory is isomorphic to the
cone ${\cal C}\left( \n010\right)$.
When the theory is promoted to an non-abelian one, there are naively
conformal operators whose existence is in contradiction with geometric
expectations and with the KK spectrum, in this case the hypermultiplets
that do not satisfy relation~(\ref{irredu}).
Differently from what happens for $T^{1,1}$ \cite{witkleb},
the superpotential which can be added to the theory is not sufficient
for eliminating these redundant non-abelian operators.
\section{Tests of correspondence}\label{comparison}
In this section we present the basic checks of the correspondence
between the ${\cal N}\!=\!3$ superconformal gauge theory just
discussed and the $\n010$ compactification of M-theory.
Here we identify the whole set of $\mathrm{BPS}$ composite operators
dual to the short supermultiplets of the $\mathrm{KK}$ spectrum.
In the next section we analyze the non-$\mathrm{BPS}$ composite operators
dual to certain massive $\mathrm{KK}$ modes, which seem to be organized
into the Higgs supermultiplet of a spontaneously broken ${\cal N}\!=\!4$
supergravity.
\subsection{Comparison with the $\mathrm{KK}$ spectrum}
Let us briefly summarize the $\mathrm{KK}$ spectrum of the
AdS$_4\times \n010$ compactification of $11D$ supergravity, organized
into ${\cal N}\!=\!3$ supermultiplets \cite{n010massspectrum,osp34}.
These are listed in table \ref{N=3massless} and \ref{N=3short},
where we give their decomposition in ${\cal N}\!=\!2$
supermultiplets and their flavor quantum numbers.
The ultrashort multiplets are:
\begin{equation}
\begin{array}{|c|c|c|}
\hline
{\cal N}\!\!=\!\!3\ {\rm multiplet}&
\left\{\begin{array}{c}
{\rm R-charge}\\
\mathrm{SU(3)}{\rm -irrep}
\end{array}\right.&
{\cal N}\!\!=\!\!2\ {\rm multiplets}\\
\hline
\hline
{\rm massless\ graviton}&
\left\{\begin{array}{c}
J=0\\
M_1=M_2=0
\end{array}\right.&
\left\{\begin{array}{c}
1\ {\rm massless\ graviton}\\
1\ {\rm massless\ gravitino}
\end{array}\right.\\
\hline
{\rm massless\ vector}&
\left\{\begin{array}{c}
J=1\\
M_1=M_2=0
\end{array}\right.&
\left\{\begin{array}{c}
1\ {\rm massless\ vector}\\
2\ {\rm chiral\ mult.}
\end{array}\right.\\
\hline
{\rm massless\ vector}&
\left\{\begin{array}{c}
J=1\\
M_1=M_2=1
\end{array}\right.&
\left\{\begin{array}{c}
1\ {\rm massless\ vector}\\
2\ {\rm chiral\ mult.}
\end{array}\right.\\
\hline
\end{array}\label{N=3massless}
\end{equation}
The short multiplets are:
\begin{equation}
\begin{array}{|c|c|c|}
\hline
{\cal N}\!\!=\!\!3\ {\rm multiplet}&
\left\{\begin{array}{c}
{\rm R-charge}\\
\mathrm{SU(3)}{\rm -irrep}
\end{array}\right.&
{\cal N}\!\!=\!\!2\ {\rm multiplets}\\
\hline
\hline
{\rm short\ graviton}&
\left\{\begin{array}{c}
J=k\geq 1\\
M_1=M_2=k
\end{array}\right.&
\left\{\begin{array}{cc}
2&{\rm short\ gravitons}\\
2k\!-\!1&{\rm long\ gravitons}\\
2&{\rm short\ gravitinos}\\
2k\!-\!1&{\rm long\ gravitinos}\\
\end{array}\right.\\
\hline
{\rm short\ gravitino}&
\left\{\begin{array}{c}
J=k+1,\ \ k\geq 0\\
M_1=k,\ M_2=k+3
\end{array}\right.&
\left\{\begin{array}{cc}
2&{\rm short\ gravitinos}\\
2k+1&{\rm long\ gravitinos}\\
2&{\rm short\ vectors}\\
2k+1&{\rm long\ vectors}\\
\end{array}\right.\\
\hline
{\rm short\ vector}&
\left\{\begin{array}{c}
J=k,\ \ k\geq 2\\
M_1=M_2=k
\end{array}\right.&
\left\{\begin{array}{cc}
2&{\rm chiral\ mult.}\\
2&{\rm short\ vectors}\\
2k\!-\!1&{\rm long\ vectors}
\end{array}\right.\\
\hline
\end{array}\label{N=3short}
\end{equation}
\subsubsection{The fundamental supersingletons}
In complete analogy with the ${\cal N}\!=\!2$ CFT's analyzed
in \cite{3dcft}, the building blocks of all the superconformal
primary fields are the supersingletons.
In this case we have at our disposal the isospin doublet in the
fundamental representation of the flavor group $\mathrm{SU(3)}$:
\begin{equation}\label{N=3supersingleton}
\Theta^{i\,J=1/2}=\left(\begin{array}{c}
\Theta^{+i}\\
\Theta^{-i}
\end{array}\right)^{{\cal N}\!=\!3}=
\left(\begin{array}{c}
U^i\\
i\overline V^i
\end{array}\right)^{{\cal N}\!=\!2}-\ft{\sqrt 2}{2}\,
\theta^0\!\left(\begin{array}{c}
i{\cal D}^+\overline V^i\\
{\cal D}^- U^i
\end{array}\right)^{{\cal N}\!=\!2}\,,\label{N=3doublet}
\end{equation}
and the conjugate doublet
\begin{equation}
\Theta_j^{\ J=1/2}=\left(\begin{array}{c}
\Theta^+_j\\
\Theta^-_j
\end{array}\right)^{{\cal N}\!=\!3}=
\left(\begin{array}{c}
-iV_j\\
\overline U_j
\end{array}\right)^{{\cal N}\!=\!2}-\ft{\sqrt 2}{2}\,
\theta^0\!\left(\begin{array}{c}
{\cal D}^+\overline U_j\\
-i{\cal D}^- V_j
\end{array}\right)^{{\cal N}\!=\!2}\,.\label{N=3conjugate}
\end{equation}
The ${\cal N}=2$ superfield components $U^i$ and $V_i$ are
supersingletons:
\begin{eqnarray}
  U^i(x,\theta^\pm) &=& u^i(x) + \theta^+\chi^{-\,i}_u(x)
  + \ft{1}{2}\theta^+\partial\!\!\!/u^i(x)\theta^-\,,\nonumber\\
  V^i(x,\theta^\pm) &=& v_i(x) + \theta^-\chi^+_{v\,i}(x)
  - \ft{1}{2}\theta^+\partial\!\!\!/v_i(x)\theta^-\,.
\label{defiuvi}
\end{eqnarray}
The lowest components, the so-called {\it Di}'s, are the scalar
fields $u^i$ and $v_i$ discussed in the previous section
and realizing the homogeneous coordinates of
$\mathbb{P}^2 \times \mathbb{P}^{2*}$.
Their color representations are given in (\ref{assegnati}) and
is the same for the {\it Rac}'s $\chi_u$ and $\chi_v$.
\subsubsection{Field theory realization of the chiral ring}
The generator of the chiral ring of our CFT is the highest
weight part of the tensor product of \eqn{N=3doublet}
times its conjugate doublet, i.e. the $\mathrm{SO(3)_R}$ triplet
\begin{equation}\label{N=3triplet}
\Theta^{i\ J=1}_{\,j}=Tr\left(\begin{array}{c}
-iU^iV_j\\
\frac{\sqrt{2}}{2}\left(U^i\overline U_j+\overline V^iV_j\right)\\
i\overline V^i\overline U_j
\end{array}\right)^{{\cal N}\!=\!2}+{\cal O}(\theta^0)\,,
\end{equation}
where the trace refers to the color indices.
From the flavor viewpoint, we can extract the two irreducible pieces
belonging to the symmetric tensor product of the $\bf 3$ and
$\bf \bar 3$ of $\mathrm{SU(3)}$.
They contain the two massless vectors in the Kaluza Klein spectrum
(see table \eqn{N=3massless}):
\begin{itemize}
\item the adjoint,
\begin{equation}
\Sigma^i_{\,\,j}\equiv\frac{\sqrt{2}}{2}
Tr\left(U^i\overline U_j+\overline V^iV_j\right)-{\rm flavor\ trace}
\end{equation}
corresponding to the conserved current of the global $\mathrm{SU(3)}$ flavor;
\item the singlet,
\begin{equation}
\Sigma\equiv\frac{\sqrt{2}}{2}
Tr\left(U^i\overline U_i+\overline V^iV_i\right)
\end{equation}
corresponding to the baryonic $\mathrm{U(1)}$ global symmetry.
\end{itemize}
\par
By composing several massless vectors we obtain the whole
chiral ring of superfields, containing ${\cal N}\!=\!2$
chiral fields and short vectors with the right flavor quantum
numbers, as listed in table \eqn{N=3massless}.
\subsubsection{Field theory realization of the short gravitinos}
Let us come to the short gravitinos.
Remember that we basically have at our disposal the ${\cal N}\!=\!3$
supersingleton of (\ref{N=3supersingleton}), which we will simply
call $\Theta^i$, and its conjugate $\Theta_j$.
Let us consider the following composite operator:
\begin{equation}
\Theta^{(ijk)}=f^{lm(i}Tr\left[\Theta^j\Theta^{k)}
\Theta_l\Theta_m\right]\,,\label{shgravino}
\end{equation}
where $f^{ijk}$ are the $\mathrm{SU(3)}$ structure constants and the round
brackets mean symmetrization.
From the isospin viewpoint, \eqn{shgravino} is a triplet, while
it transforms in the three time symmetric tensor
product of the $\bf{3}$ of $\mathrm{SU(3)}$, in agreement with the
$J\!=\!1$ short gravitino of the $\n010$
Kaluza Klein spectrum (see table \ref{N=3short}).
By construction, the operator (\ref{shgravino}) is a short gravitino,
namely it satisfies the second order differential constraint
of eq. (\ref{diffconstraint2}).
\\
The ${\cal N}\!=\!2$ superfield content (see eq. \ref{J=1shgravino})
is given by:
\begin{eqnarray}
\Sigma^{+(ijk)}=f^{lm(i}\ U^j U^{k)}
\left(V_l\overline U_m-V_m\overline U_l\right);\\
\Sigma^{\,0\,(ijk)}=\sqrt 2 if^{lm(i}\ U^j \overline V^{k)}
\left(V_l\overline U_m-V_m\overline U_l\right);\\
\Sigma^{-\,(ijk)}=-f^{lm(i}\ \overline V^j\overline V^{k)}
\left(V_l\overline U_m-V_m\overline U_l\right);\\
G^{+(ijk)}_\alpha=f^{lm(i}\ U^j U^{k)}
\left(\overline U_l {\cal D}_\alpha^+\overline U_m-\overline U_m
{\cal D}_\alpha^+\overline U_l\right);\\
G^{-(ijk)}_\alpha=-f^{lm(i}\ \overline V^j\overline V^{k)}
\left(V_l{\cal D}_\alpha^-V_m-V_m{\cal D}_\alpha^-V_l\right).
\end{eqnarray}
The ${\cal N}\!=\!3$ short gravitinos of higher isospin are
obtained by extracting the highest weight part from
the product of operators in the chiral ring with \eqn{shgravino}:
\begin{equation}
\Theta^{(i_1\cdots i_{k-1}klm)\ J=k}_{\ (j_1\cdots j_{k-1})}=
Tr\Big[\underbrace{\Theta^{i_1}_{\ j_1}\otimes_{h.w.}\cdots
\otimes_{h.w.}\Theta^{i_{k-1}}_{\ j_{k-1}}}_{k-1\ {\rm objects}}
\otimes_{h.w.}\Theta^{(klm)}\Big]\,.
\end{equation}
\subsubsection{Field theory realization of the short gravitons}
Let us now consider the composite superfield:
\begin{equation}
  \Theta^{J=0}_\alpha=Tr\left[\Theta^i\otimes{\cal D}_\alpha\otimes
  \Theta_i-\Theta_i\otimes{\cal D}_\alpha\otimes\Theta^i\right]\,,
\label{mlgraviton}
\end{equation}
where the scalar part is extracted from the isospin tensor product.
It is straightforward to show that this superfield is a short
graviton (\ref{J=0shgraviton}) by construction.
From the ${\cal N}\!=\!2$ viewpoint, it is composed by:
\begin{itemize}
  \item the massless graviton supermultiplet of the ${\cal N}\!=\!2$
  subalgebra:
  \begin{eqnarray}
    T_{(\alpha\beta)}=Tr\Big[V\partial\!\!\!/_{(\alpha\beta)}\overline V
    -\overline V\partial\!\!\!/_{(\alpha\beta)}V
    -U\partial\!\!\!/_{(\alpha\beta)}\overline U
    +\overline U\partial\!\!\!/_{(\alpha\beta)}U\nonumber\\
    +2{\cal D}^-_{(\alpha}U\,{\cal D}^+_{\beta)}\overline U
    +2{\cal D}^+_{(\alpha}\overline V\,{\cal D}^-_{\beta)} V\Big]\,;
  \end{eqnarray}
  \item the conserved current relative to the third supersymmetry
  charge, completing the ${\cal N}\!=\!3$ supersymmetry algebra:
\begin{equation}
 G_\alpha=iTr\Big[U{\cal D}^-_\alpha V-V{\cal D}^+_\alpha U+
 \overline V{\cal D}^+_\alpha\overline U
 -\overline U{\cal D}^+_\alpha\overline V\Big]\,.
\end{equation}
\end{itemize}
All together, $T_{\alpha\beta}$ and $G_{\alpha}$ constitute the
supermultiplet containing the energy-momentum tensor, the ${\cal N}\!=\!3$
supersymmetry charges and the ${\cal N}\!=\!3$ R-symmetry currents.
\par
Once again, the short gravitons of the $CFT$ are realized by
composing (\ref{mlgraviton}) with the chiral
ring operators and taking the highest weight part of isospin
and flavor quantum numbers:
\begin{equation}
\Theta^{(i_1\cdots i_k)\ J=k}_{\alpha\ (j_1\cdots j_k)}=
Tr\Big[\underbrace{\Theta^{i_1}_{\ j_1}\otimes_{h.w.}
\cdots\otimes_{h.w.}\Theta^{i_k}_{\ j_k}}_{k\ {\rm objects}}
\otimes_{h.w.}\Theta^{\ J=0}_{\alpha}\Big]\,.
\label{shgraviton}
\end{equation}
It is interesting to note that some of the ${\cal N}=2$ components
of the short gravitons (\ref{shgraviton}) of $J\geq 1$ (precisely,
the second highest helicity states) are long gravitons with the
following particular structure:
\begin{equation}
  \Phi\sim{\rm conserved\ vector\ current}\times
  {\rm chiral\ operator}\times{\rm stress-energy\ tensor}\,.
\label{rational}
\end{equation}
These long ${\cal N}=2$ multiplets have nonetheless rational conformal
dimensions, belonging to a short ${\cal N}=3$ graviton multiplet.
Furthermore, they have the same structure of some long multiplets of
rational conformal dimension identified in type IIB \cite{sergiotorino}
as well as in ${\cal N}=2$ (see eq. (6.64) of \cite{3dcft})
$M$-theory compactifications.
This suggests that the existence of such {\it rational multiplets}
in non-maximally supersymmetric AdS compactifications could be
explained by the presence of a residual form of higher supersymmetry,
possibly spontaneously broken.
This explanation is confirmed by a second feature common to all the
${\cal N}=3$ AdS$_4$ compactifications of $11D$ supergravity: the
presence of a superHiggs multiplet, that we discuss in the next section.
\section{The Universal SuperHiggs multiplet}
\label{suphigs}
Finally we consider the CFT realization of a long gravitino multiplet that
has integer conformal dimension:
\begin{equation}
  E_0 =3
\label{e0is3}
\end{equation}
and it is neutral with respect to the flavor group SU(3).
It was found in the spectrum of the AdS$_4 \times \n010$
compactification \cite{osp34} but, as we shall argue in a forthcoming
paper \cite{noinext}, it has a universal character, since it would
appear with the same quantum numbers and the same conformal dimension
(\ref{e0is3}) in any other Freund Rubin compactification of $D=11$
supergravity with ${\cal N}=3$ residual supersymmetry.
In \cite{noinext} we shall discuss its interpretation as superHiggs
multiplet in a partial supersymmetry breaking
${\cal N}=4$ to ${\cal N}=3$.
Here we want to stress its universality also from the CFT point of view.
\par
Consider the following scalar composite superfield:
\begin{equation}
{\cal SH}={\rm Tr}\big[\underbrace{\Theta_\Sigma\otimes\Theta_\Sigma\otimes
\Theta_\Sigma}_{J=0}\big]={\rm Tr}\left[\Theta_\Sigma^+\Theta_\Sigma^{\,0\,}
\Theta_\Sigma^-\right]\,,\label{superHiggs}
\end{equation}
where $\Theta_\Sigma$ is the field strength superfield, i.e. a real
$J=1$ short superfield (see eq. \ref{J=1chiral}) generalizing the linear
multiplet of ${\cal N}=2$ gauge theories:
\begin{eqnarray}
  \Theta_\Sigma=\left(\begin{array}{c}
  Y\\
  \Sigma\\
  -Y^\dagger
  \end{array}\right)+{\cal O}(\theta^0)=\hspace{8cm}\\
  \left(\begin{array}{l}
  Y+(\theta^+\chi^-)+(\theta^+\theta^+)H+\ft{1}{2}(\lambda^+\theta^0)
  +\ft{1}{2}(\theta^0\theta^+)P\\
  -\ft{1}{2}(\theta^0\theta^0)H^\dagger
  -\ft{i}{2}(\theta^0\gamma^{\mu\nu}\theta^+)F_{\mu\nu}\\
  \\
  -M+\ft{1}{2}(\lambda^+\theta^-)+\ft{1}{2}(\lambda^-\theta^+)
  +\ft{1}{2}(\theta^+\theta^-)P-\ft{i}{2}(\theta^-\gamma^{\mu\nu}\theta^+)F_{\mu\nu}\\
  +\ft{1}{2}(\theta^0\chi^-)-\ft{1}{2}(\theta^0\chi^+)+(\theta^0\theta^+)H
  -(\theta^0\theta^-)H^\dagger+\ft{1}{4}(\theta^0\theta^0)P\\
  \\
  -Y^\dagger-(\theta^-\chi^+)-(\theta^-\theta^-)H^\dagger+\ft{1}{2}(\lambda^-\theta^0)
  +\ft{1}{2}(\theta^0\theta^-)P\\
  +\ft{1}{2}(\theta^0\theta^0)H
  -\ft{i}{2}(\theta^-\gamma^{\mu\nu}\theta^0)F_{\mu\nu}
  \end{array}\right)+
  \begin{array}{c}
  {\rm derivative}\\
  {\rm terms.}\nonumber
  \end{array}
\end{eqnarray}
From eq. (\ref{superHiggs}) it is possible to identify all the field
components of the superHiggs multiplet, which turn out to be related,
through the AdS/CFT correspondence, to certain Kaluza Klein modes of
the $\n010$ compactification.
Of particular interest, for its clear geometrical interpretation, is
the scalar component of zero isospin and conformal dimension $6$.
The corresponding supergravity state is given by the {\it breathing
mode}, responsible for a uniform dilatation of the internal manifold
$X^7$.
It can be extracted by integrating the superfield ${\cal SH}$ with
respect to the Grassmann measure $d^6\theta$:
\begin{equation}
\int d^2\theta^+ d^2\theta^- d^2\theta^{\,0\,}
{\cal SH}=Tr\left[3i HH^\dagger P+\ft{1}{4}\epsilon^{\lambda\mu\nu}
\epsilon^{\rho\sigma\tau}F_{\lambda\mu}F_{\nu\rho}F_{\sigma\tau}\right]
+{\rm derivatives}.
\label{FFF}
\end{equation}
The supergravity interpretation of this field as the volume mode of
$X^7$ is a clear sign of the universality of the whole multiplet (it
does not depend on any specific characteristic of the internal manifold).
As we will show in \cite{noinext} this is true for all the components
of the multiplet.

From the CFT viewpoint, the composite operator (\ref{FFF}) is the
${\cal N}=3$ supersymmetrization of the following third power of
the gauge field strength:
$$
  \epsilon^{\lambda\mu\nu}\epsilon^{\rho\sigma\tau}
  F_{\lambda\mu}F_{\nu\rho}F_{\sigma\tau}\,,
$$
whose dimension appears to be protected by some, so far unknown,
non-renormalization theorem.
Indeed a closely similar situation appears in type IIB AdS$_5$
compactifications, where the volume mode of the internal manifold
is dual to the CFT operator $F^4$, of dimension 8, which is
known to satisfy some non-renormalization theorem.
This consideration suggests that the operator (\ref{FFF}) could
originate by the low energy expansion of an analogue of the Dirac Born
Infeld for the $M2$-brane, as well as the operator $F^4$ comes from the
$\alpha'$ expansion of the DBI Lagrangian of the $D3$-brane.
$F^4$ is indeed the operator that coupled to the background breathing
mode on the D3-brane world-volume.
\par
In this perspective, the universality of the (properly supersymmetrized)
third power of $F$ could be understood: it should be traced
back to the existence of a universal Lagrangian term for the $M2$-brane.
\par
The explicit presence of $F_{\mu\nu}$ in the previous formulae deserves
some comments.
In three dimensions, the vector multiplet is not conformal and it does
not make sense to consider it an elementary degree of freedom at the
conformal point. The only singletons in three dimensions are
hypermultiplets. Only hypermultiplets indeed appeared in the matching
of the KK spectrum with the short multiplets of conformal operators
that we discussed in the previous sections. The vector multiplet fields
in the previous equations should be regarded as expressed in terms
of the singletons at the conformal point, using the equations
of motion, for example. Alternatively, we may consider the previous
equations as operators in the three dimensional gauge theory that has
the CFT as the IR limit. The previous discussion suggests that these
operators become conformal operators at the fixed point.
%
%
\section{Conclusions and perspectives}
The identification and the study of conformal field theories dual to
AdS supergravity compactifications is not a mere exercise of
classification nor a simple test of AdS/CFT correspondence.
As it is the case for the $\n010$ solution considered in this paper, a
careful analysis of the properties of the theory, both on the CFT and
on the supergravity side, may lead to surprising discoveries.
\par
The most interesting lesson we have learned about non-maximally
compactifications of $M$-theory regards the existence of some
universal features which do not depend on the geometrical details of
the compactification manifold, but only on the degree of
supersymmetry of the solution.
\par
From the supergravity viewpoint, we find that all the massless
multiplets, related to symmetries of the theory, are always coupled
to long {\it shadow} multiplets.
Some of these can be interpreted as the massive (super)Higgs multiplets
of some spontaneously broken (super)symmetry.
This phenomenon is particularly interesting for the most general
symmetries, such as the group of AdS space-time isometries and/or
its supersymmetries.
In the ${\cal N}=3$ case, for instance, the {\it shadow} multiplet
of the massless graviton, related to the $\mathrm{Osp(3|4)}$ supergroup,
is a massive gravitino multiplet with same quantum numbers of the
first: it is a superHiggs multiplet.
Hence every ${\cal N}=3$ solution of the form AdS$_4\times X^7$,
independently from $X^7$, turns out to be the broken phase of some
not better specified ${\cal N}=4$ supergravity.
The deepest implications of this fact are analyzed in \cite{noinext}.
\par
Here we want to briefly discuss the consequences of the field theory
counterpart of this phenomenon.
The AdS/CFT prescriptions imply that the composite operators
dual to the supergravity {\it shadow} multiplets have protected
conformal dimensions.
This fact is quit surprising because they are not organized in short
multiplets, suggesting the existence of some non-trivial
non-renormalization theorem, whose investigation is left to future
speculations.
\par
Another possible development is given by the $M$-brane interpretation
of the CFT dual of the most universal {\it shadow} multiplet: the
{\it shadow} of the stress-energy tensor, i.e. the {\it breathing
mode} of the internal manifold.
Its existence is independent even from the degree of supersymmetry of
the theory, hence it must come from a universal term of the $M2$-brane
action, not directly related to the background.
The identification of such a term could shed new light on the microscopic
structure of the $M$-theory.
\paragraph{Acknowledgements}
We are grateful to Cesare Reina, Alessandro Tomasiello, Alessandro Zampa for many
important discussions on the geometry of the $\n010$ manifold at the beginning of
this work. We also acknowledge important exchanges of ideas with Sergio Ferrara
and Riccardo D'Auria.
\appendix
\section{Conventions on spinors}
In accordance with \cite{susp}, spinor indices ($\alpha,\beta,\gamma\ldots$)
are contracted \emph{from\ eight\ to\ two} and are raised and lowered with
$\epsilon_{\alpha\beta}$:
\begin{eqnarray}
\psi^\alpha\equiv\epsilon^{\alpha\beta}\psi_\beta\qquad
\epsilon_{12}=\epsilon^{21}=1\nonumber\\
\psi_\alpha\equiv\epsilon_{\alpha\beta}\psi^\beta\qquad
\epsilon_{\alpha\gamma}\epsilon^{\gamma\beta}=\delta_\alpha^\beta\nonumber\\
(\psi\phi)\equiv\psi_\alpha\phi^\alpha=-\psi^\alpha\phi_\alpha=
\phi_\alpha\psi^\alpha\equiv(\phi\psi)\,.
\end{eqnarray}
We choose the following representation of the $SO(1,2)$ Clifford algebra:
\begin{eqnarray}
\left\{\begin{array}{ccc}
\gamma^{\,0\,}&=&-i\sigma^2\\
\gamma^1&=&\sigma^3\\
\gamma^2&=&\sigma^1
\end{array}\right.\qquad
\begin{array}{c}
\eta_{\mu\nu}={\rm diag}(-++)\\
\gamma^\mu\equiv\gamma^{\mu|\alpha}_{\ \ \ \ \beta}\\
\left[\gamma^\mu,\gamma^\nu\right]=2\epsilon^{\mu\nu\rho}\gamma_\rho\,,
\end{array}
\end{eqnarray}
hence the symmetry properties of the gamma matrices are:
\begin{eqnarray}
\left\{\begin{array}{ccccc}
\gamma^{\mu|\alpha\beta}&\equiv&\epsilon^{\beta\gamma}
\gamma^{\mu|\alpha}_{\ \ \ \ \gamma}&=&\gamma^{\mu|\beta\alpha}\\
\gamma^\mu_{\ \ \alpha\beta}&\equiv&\epsilon_{\alpha\gamma}
\gamma^{\mu|\gamma}_{\ \ \ \ \beta}&=&\gamma^\mu_{\ \ \beta\alpha}
\end{array}\right.
\end{eqnarray}
so that
\begin{equation}
(\psi\gamma^\mu\phi)=-(\phi\gamma^\mu\psi)\,.
\end{equation}
Complex conjugation acts as:
\begin{equation}
(\psi^\alpha)^\dagger\equiv\bar\psi_\alpha\,,
\end{equation}
so that
\begin{equation}
(\psi\phi)^\dagger=\bar\phi\bar\psi=(\bar\psi\bar\phi),
\end{equation}
and, with our choice of gamma matrices,
\begin{equation}
(\psi\gamma^\mu\phi)^\dagger=-(\bar\phi\gamma^\mu\bar\psi)=
(\bar\psi\gamma^\mu\bar\phi)\,.
\end{equation}
The spinorial derivatives act in the following way:
\begin{equation}
\frac{\partial}{\partial\theta^i_\alpha}\theta^j_\beta=
-\frac{\partial}{\partial\theta^{i\beta}}\theta^{j\alpha}=
\delta_i^j\delta^\alpha_\beta
\end{equation}
and the supercovariant derivatives are:
\begin{eqnarray}
{\cal D}^+\equiv-\left(\frac{\partial}{\partial\theta^-}
+\ft{1}{2}\partial\!\!\!/\theta^+\right),\nonumber\\
{\cal D}^-\equiv-\left(\frac{\partial}{\partial\theta^+}
+\ft{1}{2}\partial\!\!\!/\theta^-\right),\nonumber\\
{\cal D}^{\,0\,}\equiv\left(\frac{\partial}{\partial\theta^0}
-\ft{1}{2}\partial\!\!\!/\theta^{\,0\,}\right).
\end{eqnarray}
\section{Notes on the ${\cal N}\!=\!2$ superfields}
Here we briefly review the differential constraint defining the
${\cal N}=2$ short superfield and their field decomposition, to
fix the notations adopted in the paper.
\begin{itemize}
\item {\bf The chiral superfield}\\
Identified by the constraint:
\begin{equation}
{\cal D}^+\Phi(x,\theta^{\pm})=0\,.
\end{equation}
In components is given by
\begin{eqnarray}
\Phi(x,\theta^{\pm})= z(x)+\theta^+\chi^-(x)
+(\theta^+\theta^+)H(x)\nonumber\\
+\ft{1}{2}\theta^+\gamma^\mu\theta^-\partial_\mu z(x)
+\ft{1}{4}(\theta^+\theta^+)\theta^-\gamma^\mu\partial_\mu\chi^-(x)\nonumber\\
+\ft{1}{16}(\theta^+\theta^+)(\theta^-\theta^-)\Box  z(x)\,.
\end{eqnarray}
\item {\bf The supersingleton}\\
The ${\cal N}\!=\!2$ supersingleton is defined by
\begin{equation}
\left\{\begin{array}{ccc}
{\cal D}^+\Phi_s(x,\theta^{\pm})&=&0\\
({\cal D}^-{\cal D}^-)\Phi_s(x,\theta^{\pm})&=&0\,.
\end{array}\right.
\end{equation}
In components is given by
\begin{eqnarray}
\Phi_s(x,\theta^{\pm})= z(x)+\theta^+\chi^-(x)
+\ft{1}{2}\theta^+\gamma^\mu\theta^-\partial_\mu z(x)\,,
\end{eqnarray}
where $z$ and $\chi^-$ are on-shell massless fields:
\begin{equation}
\left\{\begin{array}{c}
\Box z=0\,,\\
\partial\!\!\!/\chi^-=0\,.
\end{array}\right.
\end{equation}
\item {\bf The short gravitino}\\
The short gravitino, defined by:
\begin{equation}
{\cal D}^+_\alpha G^{+\alpha}(x,\theta^{\pm})=0\,,
\end{equation}
is given by
\begin{eqnarray}
G^{+\alpha}(x,\theta^{\pm})=\lambda_L+A\!\!\!/^+\theta^-
+A\!\!\!/^-\theta^++\phi^-\theta^+\nonumber\\
+\lambda^{+-}_T(\theta^+\theta^-)
+\ft{1}{2}(\theta^+\lambda^{+-}_T)\theta^-
+(\theta^+\theta^+)\lambda^{--}_T\nonumber\\
+(\theta^+\gamma^\mu\theta^-)\psi_\mu
+(\theta^+\theta^+)Z\!\!\!/\theta^-+{\rm derivative\ terms}\,.
\end{eqnarray}
\item {\bf The massless gravitino}\\
The massless gravitino, defined by:
\begin{equation}
{\cal D}^+_\alpha G^\alpha(x,\theta^{\pm})=
{\cal D}^-_\alpha G^\alpha(x,\theta^{\pm})=0\,,
\end{equation}
is given by
\begin{eqnarray}
\Phi^\alpha(x,\theta^{\pm})=\lambda_L+A\!\!\!/^+\theta^-
+A\!\!\!/^-\theta^+
+(\theta^+\gamma^\mu\theta^-)\psi_\mu
+{\rm derivative\ terms}\,,
\end{eqnarray}
where the spinor $\lambda_L$ and the gravitino $\psi_m$ are massless:
\begin{eqnarray}
\partial\!\!\!/\lambda_L=\epsilon^{\mu\nu\rho}\gamma_\mu\partial_\nu\psi_\rho=0\,,
\end{eqnarray}
while the two vectors are in Lorentz gauge:
\begin{equation}
\partial\cdot A^+=\partial\cdot A^-=0\,.
\end{equation}
\item {\bf The gauge potential superfield}\\
Identified by the reality constraint
\begin{equation}
V^\dagger=V\,,
\end{equation}
can be parametrized as:
\begin{eqnarray}
V(x,\theta^+,\theta^-)=C(x)+\theta^+\psi^-(x)+\theta^-\psi^+(x)\nonumber\\
+(\theta^+\theta^+)B(x)
+(\theta^-\theta^-)B^\dagger(x)\nonumber\\
-\ft{i}{2}\theta^+\gamma^\mu\theta^-A_\mu(x)
+\ft{1}{2}(\theta^+\theta^-)M(x)\nonumber\\
+\ft{1}{4}(\theta^+\theta^+)\theta^-
\left[\lambda^-(x)+\gamma^\mu\partial_\mu\psi^-(x)\right]\nonumber\\
+\ft{1}{4}(\theta^-\theta^-)\theta^+
\left[\lambda^+(x)+\gamma^\mu\partial_\mu\psi^+(x)\right]\nonumber\\
+\ft{1}{8}(\theta^+\theta^+)(\theta^-\theta^-)
\left[P(x)+\ft{1}{2}\Box C(x)\right]\,.
\end{eqnarray}
The gauge transformation
\begin{equation}
V\ \to\ V+\Phi+\Phi^\dagger\,,
\end{equation}
corresponds to
\begin{equation}
\left\{\begin{array}{ll}
C\ \to\ C+z+\bar{z} & P\ \to\ P\\
\psi^\pm\ \to\ \psi^\pm + \chi^\pm & \lambda^\pm\ \to\ \lambda^\pm\\
B\ \to\ B+H & M\ \to\ M\\
A_\mu\ \to\ A_\mu+i(\partial_\mu z-\partial_\mu\bar z)&
\end{array}\right.
\end{equation}
In Wess Zumino gauge, $V$ reduces to:
\begin{eqnarray}
V(x,\theta^+,\theta^-)=
-\ft{i}{2}\theta^+\gamma^\mu\theta^-A_\mu(x)
+\ft{1}{2}(\theta^+\theta^-)M(x)\nonumber\\
+\ft{1}{4}(\theta^+\theta^+)\theta^-\lambda^-(x)
+\ft{1}{4}(\theta^-\theta^-)\theta^+\lambda^+(x)
+\ft{1}{8}(\theta^+\theta^+)(\theta^-\theta^-)P(x)\,.
\end{eqnarray}
\item {\bf The field strength}\\
The gauge invariant super field strength is a real
{\bf linear superfield}:
\begin{equation}
{\cal D}^+{\cal D}^+\Sigma={\cal D}^-{\cal D}^-\Sigma=0\,.
\end{equation}
It is derived by the potential superfield $V$ by:
\begin{eqnarray}
\Sigma\equiv{\cal D}^+_{\alpha}{\cal D}^{-\alpha}V=
{\cal D}^-_{\alpha}{\cal D}^{+\alpha}V=\nonumber\\
-M+\ft{1}{2}(\lambda^+\theta^-)+\ft{1}{2}(\lambda^-\theta^+)
+\ft{1}{2}(\theta^-\theta^+)P
-\ft{i}{2}(\theta^-\gamma^{\mu\nu}\theta^+)F_{\mu\nu}\nonumber\\
-\ft{1}{8}(\theta^-\theta^-)\theta^+\partial\!\!\!/\lambda^+
-\ft{1}{8}(\theta^+\theta^+)\theta^-\partial\!\!\!/\lambda^-
+\ft{1}{16}(\theta^+\theta^+)(\theta^-\theta^-)\Box M\,.
\end{eqnarray}
where
\begin{equation}
F_{\mu\nu}\equiv\ft{1}{2}\left(\partial_\mu A_\nu-\partial_\nu A_\mu\right)\,.
\end{equation}
\item {\bf The SYM and CS action}\\
The abelian SYM action is:
\begin{eqnarray}
-4\int d^3x\,\Sigma^2|_{(\theta^+\theta^+)(\theta^-\theta^-)}=
-4\int d^3x\,d^2\theta^+\,d^2\theta^-\,\Sigma^2\nonumber\\
=\int d^3x\,\left\{
-\ft{1}{4}\left(\lambda^+\partial\!\!\!/\lambda^-
+\lambda^-\partial\!\!\!/\lambda^+\right)
+\ft{1}{2}P^2-\ft{1}{2}\partial^\mu M\partial_\mu M
-F_{\mu\nu}F^{\mu\nu}\right\}\,.
\end{eqnarray}
The supersymmetric generalization of the Chern Simons term is:
\begin{eqnarray}
4\int d^3x\,\Sigma V|_{(\theta^+\theta^+)(\theta^-\theta^-)}=
4\int d^3x\,d^2\theta^+\,d^2\theta^-\,\Sigma V\nonumber\\
=\int d^3x\,\left\{-\epsilon^{\mu\nu\rho}F_{\mu\nu}A_\rho
-\ft{1}{2}\lambda^+\lambda^--PM\right\}\,.
\end{eqnarray}
\end{itemize}

\end{document}